# Anomalous Resonance Frequency Shift in Liquid Crystal-Loaded THz Metamaterials

Eleni Perivolari[1,*], Vassili A. Fedotov[2], Janusz Parka[3], Malgosia Kaczmarek[1], and Vasilis Apostolopoulos[1]

1. Physics and Astronomy, University of Southampton, Highfield SO17 1BJ, UK
2. Optoelectronics Research Centre & Centre for Photonic Metamaterials, University of Southampton, Highfield SO17 1BJ, UK
3. Institute of Applied Physics, Military University of Technology, 2 Kaliskiego Str., 00-908 Warsaw, Poland

**Abstract**

**Babinet complementary patterns of a spectrally tunable metamaterial incorporating a nematic liquid crystal are normally assumed to exhibit the same tuning range. Here we show that for a hybrid, terahertz liquid crystal-metamaterial, the sensitivity of its resonances to the variations of the refractive index differs substantially for the two complementary patterns. This is due to a mismatch between the alignment of the liquid crystal and the direction of the local electric field induced in the patterns. Furthermore, and more intriguingly, our experimental data indicate that it is possible to shift the resonance of the positive metamaterial pattern beyond the limit imposed by the alignment mismatch. Our analysis suggests that the observed anomalous frequency shift result from the orientational optical nonlinearity of a nematic liquid crystal.**

Compact optical components that can efficiently control terahertz radiation (typically associated with 0.1 – 10 THz band of electromagnetic spectrum) are the focus of intense research given its significant potential for applications in security screening, sensing, imaging, non-destructive evaluation and high-speed wireless communication [1-4]. However, the lack of an appropriate response at these frequencies in naturally available materials renders the development of THz optical components challenging. As a result, engineering artificial materials, the so-called metamaterials [5], are becoming one of the mainstream solutions for the THz technology [6-8].

Electromagnetic metamaterials (MMs) have advanced rapidly and had a major impact in technologies spanning from RF and microwave technologies to photonics and nanophotonics [9]. The MMs concept not only has brought to life such exotic optical effects as artificial magnetism [10,11], negative refraction [12], and cloaking [13], but also enabled dramatic enhancement of light-matter interaction leading to amplified absorption [14], asymmetric transmission [14], giant polarization rotation [15], and slow light [16,17].

The key to the enhanced light-matter interaction in MMs is a narrowband resonant response, which can be engineered in metallic MMs at virtually any frequency within the THz band taking the advantage of very large dielectric constants of metals. Although the frequencies of MM resonances are determined by the structure of MMs and, therefore, are fixed by the design, they can be tuned with the help of functional materials integrated into the fabric of MMs. Nematic liquid crystals (NLCs) were arguably the first functional media successfully exploited for active control of MMs [18,19], offering an easy-to-implement control mechanism based on reversible refractive index change [20]. LC-loaded MMs have since become a very popular and well-researched artificial material system with many applications across the electromagnetic spectrum, and in particular, in the THz domain (see, for example, [21] and references therein).

In this letter we show that such a hybrid material system still holds some surprising and unexplored effects. In particular, we report that THz metallic MMs with Babinet complementary patterns do not necessarily exhibit the same frequency tuning range when integrated with an NLC. We also show that the extent of the difference is controlled by two opposing effects: (i) mismatch between LC alignment and the direction of local electric fields induced in the MMs, and (ii) orientational (Kerr) optical nonlinearity of NLCs. We discover that, remarkably, integration of NLCs with metallic MMs enables the enhancement of such nonlinearity to the level that it can be engaged using low-intensity experiments in the THz domain.

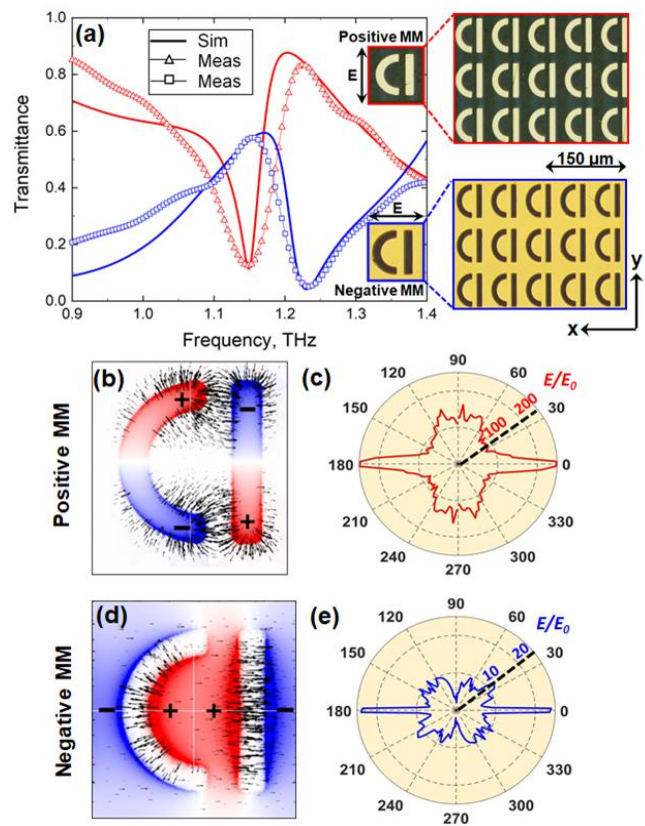

**Figure 1: (a)** Transmission spectra measured experimentally (open triangles) and simulated (thick solid lines) for pristine MMs of positive (red) and negative (blue) designs. **Insets:** Images of the unit cells and fragments of the fabricated MMs of positive (top) and negative (bottom) designs taken with a reflection microscope. **(b)** Normalised distribution of the surface charge density (colours) and electric field lines (black arrows) induced in the positive MM at the resonant frequency. **(c)** Polar plot of in-plane directivity of the local electric field in positive MM averaged over the area of the unit cell. The radius corresponds to the magnitude of the average electric field (E) normalised on the incident field ($E_0$). **(d)** Same as (b) but for negative MM. **(e)** Same as (c) but for negative MM.

The design of MMs we chose for our study was based on a D-shaped metamolecule (see insets to Figure 1a) – a derivative of the asymmetrically-split ring resonator, which in the past had enabled the engineering of high-Q Fano resonances in MMs [16] and served as a building block for one of the first MM-based sensing platforms [22]. The two Babinet complementary patterns of the MMs featured arrays of such metamolecules formed, respectively, by metal patches (positive MM) and slits in a metal screen (negative MM), as shown in the inset to Figure 1a. The width of the patches/slits was 10 µm, and the unit cell of the arrays had the dimensions of 70 µm × 80 µm. Figure 1a shows the transmission spectra of both MMs simulated in COMSOL 5.5 Multiphysics for normally incident linearly polarised plane waves, assuming that the patterns were cut in a 300 nm thick gold film [23] and supported by a 100 µm thick slab of fused quartz with the refractive index of 2 [24]. The spectra reveal the presence of Fano resonances for both positive and negative MMs, featuring the characteristic asymmetric profile [25,26] with a sharp roll-off centred at around 1.17 THz and 1.22 THz, respectively. To engage Fano resonances the polarization was set parallel to the straight segments of the metamolecules for positive MM (y-polarisation), and perpendicular to the straight segments of the metamolecules for negative MM (x-polarisation), as illustrated in the insets to Figure 1a. The transmission spectra simulated for the orthogonal polarizations were featureless, as expected.

Both resonances result from the excitation of the so-called trapped modes [16], where the charge oscillations induced in the two segments of the D-shaped metamolecule occur in anti-phase (see Figures 1b and 1d). We note that the corresponding distributions of the electric field differ substantially for the positive and negative MMs. The 'hotspots' of the electric field appear to localise around the ends of the metal patches in the positive MM, and within the central areas of the slits in the negative MM. As a result, the extent of the in-plane divergence of the local field in the two cases is also very different. Indeed, in the positive MM the field lines are seen to fan out within the hotspots, spanning nearly all directions (see Figure 1b), while in the negative MM they stretch across the slits and, thus, align predominantly along the *x*-axis (see Figure 1d). The difference becomes more apparent when analysed in terms of the directivity of the local field averaged over the area of a unit cell (see diagrams in Figures 1c and 1e). Clearly, in the case of the negative MM the directivity collapses around the x-axis and is only perturbed by four small diagonal side lobes generated by the curved segment of the metamolecule. In the case of the positive MM, however, the directivity features four broad main lobes oriented along the x- and y-axes. We argue that such a directional anisotropy of the local field, which arises near the surface of the MMs becomes an important factor in the presence of optically anisotropic materials (such as NLCs), affecting differently the tuneability of Babinet complementary MMs.

To the best of our knowledge such an effect has not been discussed in the literature. To demonstrate it first numerically, we introduced in our model a layer of an optically anisotropic dielectric, which represented an NLC in the planar state. It had the thickness of 20 µm, typical for conventional LC optical cells, and was placed on top of the MMs, fully encompassing their structure (see Appendix, Figure A1). The ordinary and extraordinary refractive indices of an NLC were assumed to be $n_\mathrm{o} = 1.574 - i0.017$ and $n_\mathrm{e} = 1.951 - i0.024$, corresponding to the refractive indices of highly birefringent LC 1825 at 1 THz [27]. Figures 2b and 2c compare the simulated transmission spectra of the complementary MMs with two different planar configurations of the NLC aligned, respectively, parallel (planar 1) and perpendicular (planar 2) to the incident polarisation. Evidently, upon switching the NLC between planar 1 and planar 2 configurations, the resonance of the positive MM exhibits a spectral shift some 20 GHz (~ 50%) smaller than that of the negative MM. This is the manifestation of the effect of the mismatch between the direction of the local field and LC alignment, which we referred to above. Here, the tendency of the local electric field to oscillate along two orthogonal directions in the positive MM (see diagram in Figure 1b) makes the supported electromagnetic mode less sensitive to alignment of the NLC (and, correspondingly, anisotropy of its refractive index) and therefore results in a spectrally narrower tuning range.

We demonstrate the extent of the effect in Figure 2a by comparing the maximum frequency shifts exhibited by resonances of the complementary MMs for different values of NLC birefringence, $\Delta n = n_e - n_o$. In our modelling $\Delta n$ varied in the range from 0.2 to 0.38, where the limits correspond to the birefringence of two different liquid crystals, E7 [28] and LC 1825 [27,29] at 1 THz. The values of $\Delta n$ within this range were calculated based on $n_e$ and $n_o$ interpolated linearly between those of E7 and LC 1825. The data points in Figure 2a mark the difference between the resonance frequencies calculated for planar 1 and planar 2 states. As expected, the frequency shift becomes smaller for both MMs as the value of the birefringence decreases towards 0.2, yet the positive MM exhibits consistently smaller shifts than the negative MM for the entire range of $\Delta n$.

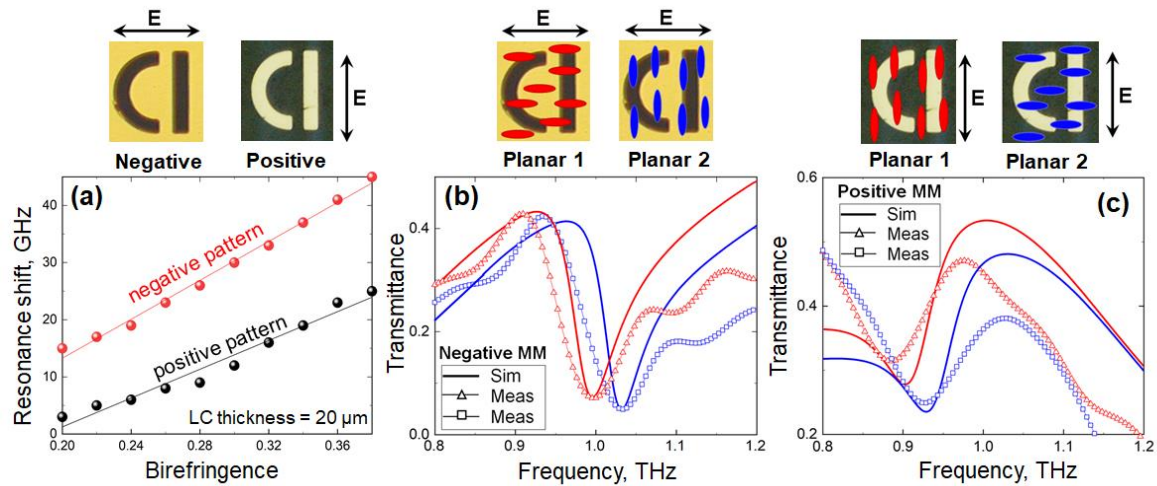

**Figure 2: (a)** Calculated maximum resonance shift that can be attained in positive (black points) and negative (red points) MMs at different values of LC birefringence. **Insets:** Images of MM unit cells taken with a reflection microscope. **(b)** Transmission spectra measured experimentally (open triangles) and simulated (thick solid lines) for negative MM loaded with NLC in planar 1 (red) and planar 2 (blue) configurations. **Insets:** Orientations of LC molecules schematically shown for planar 1 and planar 2 configurations. **(c)** Same as (b) but for positive MM.

The predicted difference in the tuneability of Babinet complementary MMs represents a practically important limitation. To confirm it experimentally, we fabricated the exact copies of the above MMs and characterised their spectral response at normal incidence using a conventional low-power THz-TDS setup, which featured a micro-dipole based photoconductive antenna excited with 10 mW of ultrafast (100 fs) 780 nm laser and measurement based on the Pockels effect. The patterns of the MMs were etched using UV photolithography in a 305 nm thick metal film, which had been deposited beforehand by thermal evaporation onto 1.17 mm thick fused quartz substrate. The film had a 300 nm thick layer of gold and a 5 nm thick layer of chromium added to ensured adhesion between gold and quartz. The overall size of the fabricated samples was 16 mm × 16 mm. The terahertz beam was focussed onto the samples to a 5 mm large spot, which guaranteed the absence of beam clipping and diffraction at the edges of the samples upon illumination. The measured transmission spectra of the pristine positive and negative MMs are shown in Figure 1a. The plot reveals a very good agreement between the experimental and modelled data confirming high quality of the fabricated MM samples.

Each MM was functionalised with LC 1825 via the planar cell arrangement, where a 20 μm thick layer of the NLC was sandwiched between the MM and a pristine slab of fused quartz (1.17 mm thick), as schematically shown in figure A1. The surface of the MM and the surface of the quartz slab facing the NLC were both coated with a thin film (~30 nm) of uniformly rubbed polyimide (PI-2525 from HD MicroSystems). Such a film acted as an alignment layer, which promoted the orientation of LC molecules near its surface in the direction of rubbing. The planar alignment of LC 1825 in the bulk was ensured by matching the directions of rubbing at the opposite sides of the resulting optical cell. To minimise fabrication inconsistencies here, the two variants of the planar alignment, namely planar 1 and planar 2, were produced simultaneously in different parts of the same cell.

The transmission spectra of LC-loaded MMs, as measured with our THz-TDS setup, are presented in Figures 2b and 2c. Clearly, the locations of the resonance exhibited by the negative MM in the presence of differently aligned NLC (i.e., in planar 1 and planar 2 configurations) agree very well with the predictions of our model, with the spectral separation reaching the expected 45 GHz. In the case of the positive MM, however, a good spectral overlap between the experimental and modelled data is seen only for the NLC in planar 2 configuration. Surprisingly, switching the NLC to planar 1 configuration red-shifts the measured transmission spectrum of the positive MM by about 20 GHz further than what seems to be allowed from the simulation. As a result, the frequency tuning range obtained experimentally for the positive MM widens to 43 GHz, which practically negates the large difference in the expected tuneability of Babinet complementary MMs which has been expected by the simulation.

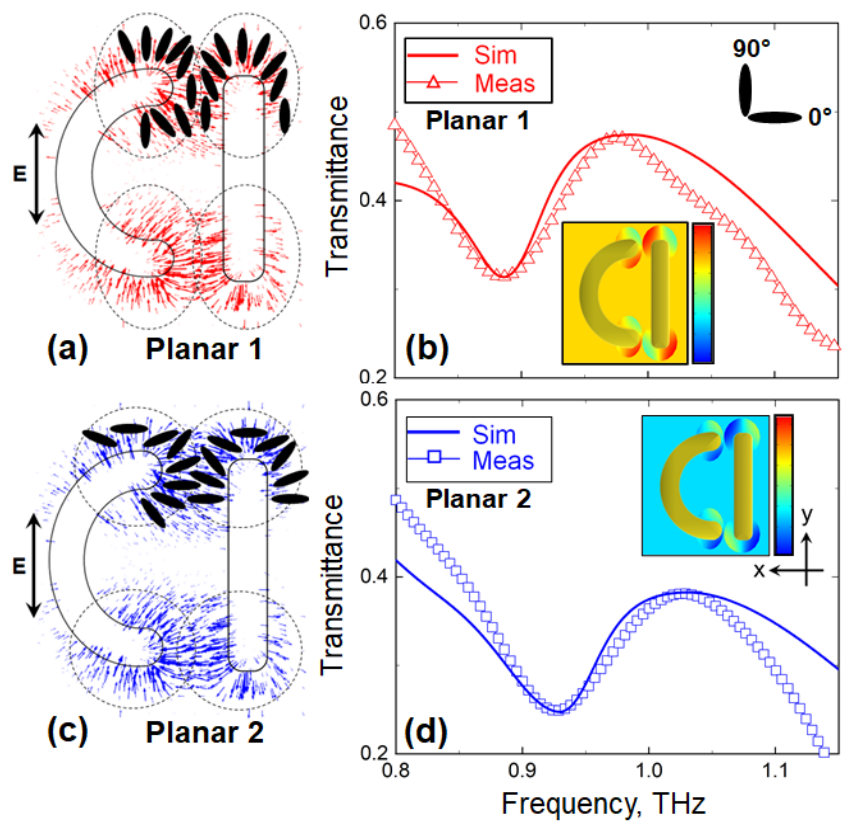

**Figure 3: (a)** Orientational optical nonlinearity of NLC in planar 1 configuration engaged with positive MM. Local field-induced distortions of the initial alignment assumed within the hotspots are shown schematically via the orientation of LC molecules. **(b)** Transmission spectra measured experimentally (open triangles) and simulated (thick solid lines) for positive MM taking into account the effect of orientation nonlinearity of NLC in planar 1 configuration. **Inset:** Spatial distribution of LC alignment assumed in the simulations. Scale bar ranges from -45° (blue) to 135° (red). 0° and 90° correspond to LC molecules being parallel to x- and y-axes, respectively. **(c)** Same as (a) but for initial planar 2 alignment of NLC. **(d)** Same as (d) but for initial planar 2 alignment of NLC.

We reason that the larger red-shift of the MM spectrum could have resulted only from a better match between the direction of the local electric field and the orientation of LC molecules, since that would effectively increase the refractive index of the NLC perceived by the resonant electromagnetic mode of the metamolecules. To this end we note that the directional spread of the local field in the positive MM is largest around the y-axis (see diagram in Figure 1b) and, hence, matching the spread with local distortions of the NLC will have the biggest impact for planar 1 alignment – exactly where the outcomes of our simulations and experiment disagree the most.

To demonstrate the plausibility of our hypothesis we re-modelled the response of the positive MM assuming that the NLC aligned itself in the direction of the local field within the hotspots (where the divergence of the field is strongest), and such distortions were confined to a distance of 5 µm away from the edges of the metamolecules and 1 µm above their plane (see Appendix, Figure A2). We further assumed that changes in LC alignment there were limited to (-45°) – (+45°) range of angles, which corresponded to the angular spread of the local field we noted above. For the directions outside the spread the orientation of LC molecules was gradually restored to the background alignment (i.e., planar 1 configuration), as schematically illustrated in Figures 3a and 3c. The transmission spectrum of the positive MM calculated using our modified model is presented in Figure 3b, where it is compared with the spectrum measured experimentally. Evidently, this time the predicted and the actual locations of the MM resonance coincide. Moreover, a very good agreement between the simulated and experimental data is now seen in terms of the amplitude of the MM transmission as well.

As a further test of our hypothesis we also modified the model of the positive MM with the NLC featuring planar 2 alignment. Localised distortions of the alignment were introduced in the model in the same way as in the planar 1 case except that the re-orientation of LC molecules was traced relative to the x-axis (see Figure 3d). What we found was that even in the planar 2 case (where the theory and experiment had agreed qualitatively) assuming re-alignment of the NLC within the hotspots dramatically improved the accuracy of our model and enabled us to achieve nearly perfect quantitative agreement between the calculated and measured spectra (see Figure 3d). For clarity different angular re-alignment ranges are simulated for both planar 1 and planar 2 cases to match the experimental data. Evidently, re-orientation of ± 45° (90°±45° and 0°±45°, respectively) found to be closer to the experiment as shown in Appendix, Figure A3.

Given that the re-alignment of the NLC must occur within the hotspots in simultaneously all metamolecules upon illumination of the sample, it could not be a spontaneous process. We therefore conclude that it was induced by the electric fields of the metamolecules and so the anomalous resonance frequency shift observed experimentally in the positive MM was the manifestation of Kerr (orientational) optical nonlinearity of the NLC [30].

To the best of our knowledge, Kerr nonlinearity of LCs has not been reported for the intensities accessible with conventional commercial THz-TDS setups (typically built around mW lasers pumping photoconductive emitters) and we argue that the following factors could have enabled the effect in our experiments. We note firstly that the re-orientation of the NLC in the configuration described here is equivalent to in-plane electrical switching, which would occur without a threshold for the initial misalignment of the NLC of less than 45°, as follows from [33]. Furthermore the E-field enhancement in our structures is dominated by two combined effects. One effect is the amplification of THz fields by the MMs via their high-Q resonant response. The other effect is the concentration of electric fields by sharp geometric features of the metamolecules, the so-called 'lightning rod' effect [31,32]. The first (resonant) effect is relevant to both the negative and positive MMs and allows local fields one order of magnitude stronger than the incident field (as evident from diagrams in Figures 1c,e). The second (geometrical) effect, however, ensures further ten-fold enhancement only in the positive MM, where sharp geometric features of the metamolecules and localisations of the resonantly amplified fields (i.e., hotspots) coincide. As a result, the overall field enhancement attainable in the positive MM is likely to exceed two orders of magnitude (see diagram in Figure 1c). We, therefore, argue that the high E-field enhancement in combination with the threshold-less nature of the LC re-orientation make it not unreasonable to expect that the enhancement of the local field in the positive MM would enable optically-induced local switching of the NLC even at the intensity levels characteristic of our THz-TDS setup.

In conclusion, we demonstrate that Babinet complementary patterns of a THz metallic MM do not exhibit the same frequency tuning range when hybridized with an NLC. The results of our study suggest that the difference results from a mismatch between the alignment of the NLC and the direction of the local electric field induced in the patterns. More intriguingly, our experimental observations indicate that it is possible to shift the resonance of the positive MM pattern beyond the limit imposed by the alignment mismatch and to significantly increase its tuning range. Our analysis suggests that this observed anomalous frequency shift results from the orientational optical nonlinearity of the NLC enhanced via integration with the metallic MM. We envisage that our findings can directly lead to the increase of the efficacy of THz modulators and other active optical components exploiting the enhanced nonlinear light-matter interactions in LC-MM hybrid structures. Most importantly, we show that nonlinear effects can be engaged with a low power photoconductive antenna-based THz spectrometer, when combined with resonant amplification and sub-wavelength concentration of THz fields facilitated by MMs. This demonstration enables future studies of nonlinear effects in the THz range, which up to now, where exclusively reserved for high-power THz spectrometers based on complex and often cumbersome, amplified laser systems.

## Acknowledgements

The authors acknowledge the financial support of the UK Engineering and Physical Sciences Research Council through grant EP/R024421/1 and the Military University of Technology of Warsaw grant for "New crystalline and composite materials for optics and photonics", UGB 521-9000-00-000, fund for the year 2020.

# Appendix

## 1. LC-loaded MM hybrid cell

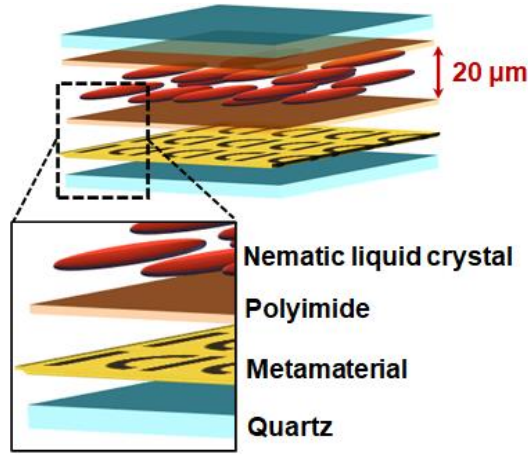

**Figure A1.** Schematic of LC-loaded MM hybrid cell used in the experiment. MM was functionalised with NLC 1825 via the planar cell arrangement, where a 20 μm thick layer of the NLC was sandwiched between the MM and a pristine slab of fused quartz (1.17 mm thick). The surface of the MM and the surface of the quartz slab facing the NLC were both coated with a thin film (~30 nm) of uniformly rubbed polyimide (PI-2525 from HD MicroSystems). Such a film acted as an alignment layer, which promoted the orientation of LC molecules near its surface in the direction of rubbing.

## 2. Modelling the change of LC birefringence induced by local fields within 'hotspots'

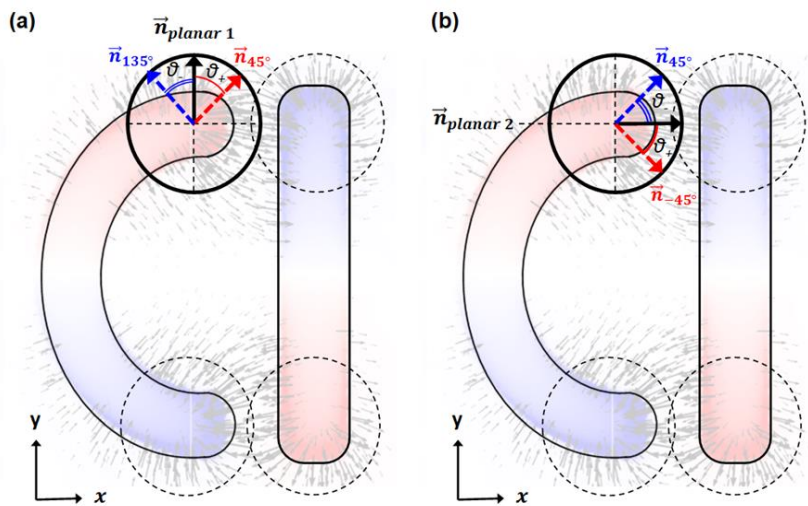

**Figure A2.** Schematic representation of modelling the anisotropy induced to the NLC molecules in **(a)** planar 1 and **(b)** planar 2 configurations due to the charge density distribution at the surface of positive MM patterns.

The hotspots are localised at the ends of the metal patches in the positive MM, as schematically shown in Figure A2. The layer of the NLC covering the MM was assumed to be aligned along the y-axis, except in the hotspots, where maximum localisation of electric field occurs. Birefringent materials such as the NLC experience double refraction, whereby light is split, depending on the incident polarization, into two, i.e. ordinary ($n_o$) and extraordinary ($n_e$) polarizations. The extraordinary one is parallel to the LC director, $\vec{n}$, of NLC molecules, pointing along the y-axis, which is the resonant polarization. This leaves the ordinary to be parallel to the x-axis direction. In order to model such optical behaviour, a diagonal optical anisotropy must be introduced in our calculations as well as a full tensor relative permittivity approach. Here, tensor refers to a 3-by-3 matrix that has both diagonal ($\epsilon_{xx}, \epsilon_{yy}, \epsilon_{zz}$) and off-diagonal ($\epsilon_{xy}, \epsilon_{xz}, \epsilon_{yx}, \epsilon_{yz}, \epsilon_{zx}, \epsilon_{zy}$) terms, as shown below.

$$\epsilon = \begin{bmatrix} \epsilon_{xx} & \epsilon_{xy} & \epsilon_{xz} \\ \epsilon_{yx} & \epsilon_{yy} & \epsilon_{yz} \\ \epsilon_{zx} & \epsilon_{zy} & \epsilon_{zz} \end{bmatrix}$$

NLCs belong to the uniaxial crystals' category, where only the diagonal elements of permittivity tensor are nonzero. This means that $\epsilon_{xx} = \epsilon_{zz} = n_o^2 - k_o^2 + 2in_o k_o$ and $\epsilon_{yy} = n_e^2 - k_e^2 + 2in_e k_e$. The material used in this study, LC1825, was fully investigated by Chodorow, U., et al. [27], who provided the following characteristics which we used in the simulations:

| LC1825 | 1.5 THz |
|---|---|
| $n_o$ | 1.574 |
| $n_e$ | 1.951 |
| $k_o$ | 7 [1/cm] |
| $k_e$ | 12 [1/cm] |
| $\Delta n$ | 0.38 |

So, for planar 1 configuration we introduced the following relative permittivity tensor with diagonal elements.

$$\epsilon = \begin{bmatrix} \epsilon_{xx} & 0 & 0 \\ 0 & \epsilon_{yy} & 0 \\ 0 & 0 & \epsilon_{zz} \end{bmatrix}$$

To model the spatial variation of optical anisotropy as specified in figure 3 (a, c), we introduced an off-diagonal transverse anisotropy in the XY plane. When the $\vec{n}$ of NLC lies and rotates in the XY plane, making an angle of $\vartheta$ with the x-axis (see Figure A2), the diagonal components ($\epsilon_{xx}, \epsilon_{yy}, \epsilon_{zz}$) and off-diagonal components ($\epsilon_{xy}, \epsilon_{yx}$) are nonzero.

$$\epsilon = \begin{bmatrix} \epsilon_{xx} & \epsilon_{xy} & 0 \\ \epsilon_{yx} & \epsilon_{yy} & 0 \\ 0 & 0 & \epsilon_{zz} \end{bmatrix}$$

In that case, $\epsilon_{yy}$ is governed by the extraordinary refractive index, because $\vec{n}$ lies along the principal y-axis, while $\epsilon_{xx}$ and $\epsilon_{zz}$ are governed by the ordinary refractive index. The off-diagonal elements $\epsilon_{xy}, \epsilon_{yx}$ are derived from the multiplication of the matrices as stated below.

$$\epsilon = \begin{bmatrix} \cos(\vartheta) & -\sin(\vartheta) & 0 \\ \sin(\vartheta) & \cos(\vartheta) & 0 \\ 0 & 0 & 1 \end{bmatrix} \begin{bmatrix} \epsilon_{xx} & 0 & 0 \\ 0 & \epsilon_{yy} & 0 \\ 0 & 0 & \epsilon_{zz} \end{bmatrix} \begin{bmatrix} \cos(\vartheta) & \sin(\vartheta) & 0 \\ -\sin(\vartheta) & \cos(\vartheta) & 0 \\ 0 & 0 & 1 \end{bmatrix}$$

$$= \begin{bmatrix} (\epsilon_{xx})\cos^2(\vartheta) + (\epsilon_{yy})\sin^2(\vartheta) & (\epsilon_{xx})\sin(\vartheta)\cos(\vartheta) - (\epsilon_{yy})\sin(\vartheta)\cos(\vartheta) & 0 \\ (\epsilon_{xx})\sin(\vartheta)\cos(\vartheta) - (\epsilon_{yy})\sin(\vartheta)\cos(\vartheta) & (\epsilon_{yy})\cos^2(\vartheta) + (\epsilon_{xx})\sin^2(\vartheta) & 0 \\ 0 & 0 & \epsilon_{zz} \end{bmatrix}$$

To model the distortions of LC alignment within the hotspots, we isolated the areas around the edges of MM and built the tensor for each area individually. For convenience, we set a new origin $(x_0, y_0)$ in each area as a point of reference to rotate the molecules' alignment, and hence the optical anisotropy, with respect to spatial coordinates.

Rotation angle $\vartheta$, will then be:

$$\hat{\vartheta} = \hat{\varphi} - \tan^{-1}[(x - x_0)/(y - y_0)],$$

The resulting rotating angle $\vartheta$ varies from 135 to 45 degrees for planar 1 configuration (see Fig. 3b) and from 45 to −45 degrees for planar 2 configuration (see Fig. 3d), showing a realistic idea of what we expected, as schematically shown in figure A2 (a, b) respectively. Angle $\varphi$, defines whether the variation axis symmetry exists at 90° along $n_{\text{planar 1}}$ or at 0° along $n_{\text{planar 2}}$.

## 3. Varying the extent of LC re-alignment within the hotspots

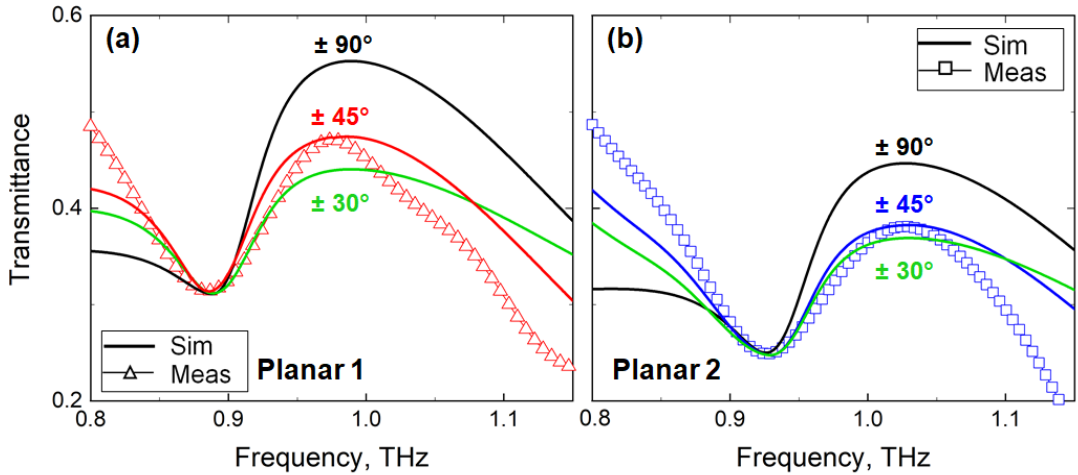

**Figure A3.** Transmission spectra of measured experimentally (open triangles) and simulated (thick solid lines) for LC-loaded positive MM in **(a)** planar 1 and **(b)** planar 2 configurations. Here, different angular re-alignment ranges are simulated for both planar 1 and planar 2 configurations to match the experimental data. Evidently, re-orientation of ± 45° (90°±45° and 0°±45°, respectively) found to be closer to the experiment. A re-orientation of ± 90° opposes the initial strong surface anchoring energy of the rubbing and hence, it seems unnatural to obtain a re-orientation of such range. Indeed, as it is shown in figure A3 re-orientation of ± 90° possesses larger transmittance than the experimental data. On the other hand, a smaller re-orientation range of ± 30° shows a weaker response. We thus conclude that the best fit is derived from the re-orientation range of ± 45°. For more details see main text.